\documentclass[12pt]{article}
\usepackage[letterpaper,top=0.75in, bottom=1in, left=1in, right=1in]{geometry}
\usepackage{graphicx}  %images
\usepackage{amsmath} %math
\usepackage{wrapfig}
\usepackage[font={small}]{caption}
\usepackage{sectsty}
\usepackage{enumitem} % to allow lists with letters instead of numbers
\usepackage{indentfirst}
\usepackage{titlesec}
\usepackage{soul} % for underlining
\usepackage[normalem]{ulem}
\usepackage[none]{hyphenat} % stop hyphenation
\usepackage{hyperref} % include urls
\usepackage{multicol} % create multi-column text
\usepackage{parskip} % to include \parindent

\titlespacing\section{0pt}{8pt plus 4pt minus 2pt}{0pt plus 2pt minus 2pt}
\sectionfont{\fontsize{12.5}{12}\selectfont}

\setuldepth{honor}

\hypersetup{%
  colorlinks=false,% hyperlinks will be black
  linkbordercolor= 1 1 1, % no box around links
  urlbordercolor= 1 1 1, % no box around links
}
\begin{document}

%%%%%%%%%%%%%%%%%%%%%%%%%%%%%%%
%%%%%%%%%%%%%%%%%%%%%%%%%%%%%%%

\begin{center}
{\bf  Isabelle Stone: breaking the glass ceiling with thin films and teaching} \\
\vspace{2mm}
Melia E. Bonomo\footnote{mbonomo@rice.edu} \\
\emph{Department of Physics \& Astronomy, Rice University, Houston, TX}
\end{center}
\vspace{3mm}

\vspace{5mm}
Dr.\ Isabelle Stone is listed as one of over four thousand \emph{American Men of Science} in the inaugural bibliographical directory published in 1906~\cite{cattell1906}.  The directory's preface states that it was compiled to recognize contributions to the advancement of pure science and to acquaint the isolated scientific man with those who have common interests so that he may be encouraged to collaborate~\cite{cattell1906}. \\

At the time of the directory's publication, Stone was a physics professor at Vassar College for women in Poughkeepsie, NY~\cite{cattell1906}.  She is not celebrated in any physics textbooks, no experiments or theories bare her name, and her research accomplishments were not Nobel-worthy.  Her teaching record was also nothing extraordinary and included several positions in secondary education schools.  At a time when physicists like Albert Einstein, Niels Bohr, and Erwin Schrodinger were making prominent contributions to the field, Stone's story is seemingly average and certainly not widely known. And yet, it is worth telling. \\

Hers is the story of forging ahead in a men's profession, having been credited as the first woman in the United States to obtain a Ph.D. in physics, one of two women founders of the American Physical Society, and one of two women to attend the first International Congress of Physics~\cite{lustig2000,creese2000,staley2008}.  Hers is the story of an unmarried, educated woman doing scientific research, living and working abroad, and providing educational stepping stones for more women to go to college in the early 1900's. It starts with the electric resistance of thin films in Chicago and becomes a journey in teaching women up and down the east coast, across the Atlantic, and back again.

%%% SECTION %%%
\vspace{5mm}
\section*{Michelson's Student}
\vspace{3mm}

Stone was born in Chicago, Illinois on October 18, though there is some discrepancy as to whether the year was 1868~\cite{cattell1906} or 1870~\cite{creese2000}.  Her parents, both originally school teachers, were prominent figures in Chicago's community---her father as an active member of the Baptist church and editor for the \emph{Chicago Times}, and her mother as an early leader in the Young Women's Christian Association~\cite{currey1918,addams2019}.  Stone followed in the footsteps of her older sister and completed a bachelor's degree in music at the all-female Wellesley College in 1890~\cite{cattell1906,addams2019}.  She then went to the University of Chicago to obtain a master's degree in 1896~\cite{cattell1906}. \\

As she studied further at the University of Chicago to obtain her Ph.D. in physics, her advisor was future Nobel laureate Albert A. Michelson.  About a decade prior, in 1887, he had worked with Edward W. Morely to disprove the existence of ether~\cite{michelson1887}.  Michelson apparently was not too keen on the idea of having graduate students, according to a memoir written by another future Nobel laureate, Robert A. Millikan~\cite{millikan1938}. Millikan overlapped with Stone in Michelson's lab in 1896 when he spent the summer working as a research scientist.  \\

Millikan recalled a conversation in which Michelson lamented over the fact that his graduate students took his project ideas and either handled the project poorly such that Michelson could not remedy it, or handled the project so well that the student thought the project was entirely their own~\cite{millikan1938}.  Between 1894 and 1904, Michelson assigned very few thesis projects, and most of those that he did assign did not turn out well~\cite{millikan1938}.  Having carried out a successful project and written her gratitude for Michelson's guidance~\cite{stone1897}, one could assume that Stone was one of the few graduate students Michelson considered a success.\\

Stone's project was the first to measure the specific conductivity of thin films~\cite{stone1898}.  She fabricated the films by depositing silver on 8 x 0.8 cm glass plates with copper terminals and a paper cushioning to avoid damaging the glass and film.  She investigated how the electrical resistance of the thin film changed over time due to the effects of age, moisture, and heat.  There was a fast drop in the resistance at short times and a much slower decrease at longer times.  In general, the higher the resistance of the film, the more rapidly that resistance decayed.  Furthermore, she found that the ratio of the observed electrical resistance to the resistance calculated from the thin film's weight, density, and dimensions was very large.  In 1897, Stone defended this work for her thesis~\cite{stone1897} and graduated as the first American woman to be awarded a physics doctorate. \\

\section*{Pioneer in the Formation of a Physics Community}
\vspace{3mm}

While still a student herself at the University of Chicago, Stone had began her career as a physics teacher and educator in general. From 1890 onward, she spent her free time teaching classes on astronomy, algebra, geometry, and women's gymnastics at Huff-House, which was a secular settlement house that provided a residence, education, and cultural center for Chicago-area immigrants~\cite{addams2019}.  Seeing as her sister also taught classes at Huff-House~\cite{addams2019} and her parents were such prominent figures in local religious organizations~\cite{currey1918}, Stone's participation may have been influenced by a deep familial value of community engagement. \\

From 1898 to 1906, Stone was an instructor in the physics department at Vassar College in Poughkeepsie, New York~\cite{leonard1914}. The women's school had been founded in 1861 and had only just formed its physics department in 1894, so Stone was an integral part of its growth.  In addition to being an instructor, Stone conducted research at Columbia University in New York City~\cite{leonard1914,addams2019} on the properties of thin films when deposited in a vacuum and on color in platinum films~\cite{stone1905}.  Outside the classroom and lab, she was an expert reader for the physics section of the College Entrance Examination Board in June 1901, which was the first time the standardized admissions exam was given~\cite{cattell1906,leonard1914}. \\

Meanwhile, the physics community was gathering steam at home and abroad.  In the U.S., there was a need for a central organization to ``advance and diffuse the knowledge of physics''~\cite{lustig2000}.  On May 30, 1899, 36 American physicists met at Columbia University to found the American Physical Society~\cite{lustig2000}.  Of note in attendance were Michelson, who became the first vice president, and two women -- Marcia Keith, who only had a bachelor's degree but was the current head of the physics department at Mount Holyoke College, and Isabelle Stone, Ph.D.~\cite{lustig2000}.  Not long after, the first International Congress of Physics, \emph{Congr\`{e}s international de physique}, was held in Paris, France in 1900 with 836 attendees~\cite{staley2008}. Once again Stone was one of only two women in attendance -- the other being Marie Curie~\cite{staley2008}.

%%% SECTION %%%
\vspace{5mm}
\section*{Spearhead of an Unconventional Education Abroad}
\vspace{3mm}

Given an increase in women's colleges in the early 1900's, the main goal of private schools for girls was to prepare them for higher education~\cite{sargent1915}.  Before becoming an instructor at Vassar College in 1898, Stone had taught at Bryn Mawr Preparatory School in Baltimore, Maryland for a year~\cite{cattell1906}, and it is possible that this experience gave her the idea to open a private girls school of her own.  \\

In 1907, she did just that with her sister, who had currently been teaching at Forest Park University in St. Louis, Missouri~\cite{leonard1914}.  Both women were entirely independent, given they were both unmarried and without children, and decided to leave their university jobs to found a private school for girls in Rome, Italy~\cite{leonard1914}. (As an interesting side note, of the 23 women physicists listed in the first three editions of \emph{American Men of Science}, none were married~\cite{rossiter1982}.) ``The Misses Stone's School for Girls,'' as they called it, was one of the few, if not only, international schools for American girls at the time~\cite{sargent1915}.  \\

The school offered a language immersion program in French and Italian
history, art history, literature, and music \cite{sargent1915}.  There were weekly supplemental lessons held around the city to experience historical sites and art galleries that the girls learned about in the classroom~\cite{sargent1915}.  Stone and her sister organized trips during the Christmas and Easter vacations to give the students opportunities to travel to neighboring regions in Italy, like Tuscany, Umbria, Campania, and Sicily, and other countries, like Spain and Greece. \\

For seven years, Stone and her sister would return to the U.S. each summer and sail with the group of girls to Italy in the fall~\cite{currey1918}.  They ran the school until the start of World War I in 1914, which forced them to close it and return to the U.S.~\cite{creese2000}.

\vspace{8pt}

%%% SECTION %%%
\section*{A Woman Academic After World War I}
\vspace{3mm}

Though men made up a much larger percentage of American scientists before 1920, over 95\%, women scientists were more educated, with 63\% having Ph.D. degrees as opposed to only 47\% for men~\cite{rossiter1974}.  In general the women were less visible and had less access to research facilities~\cite{rossiter1974}. Though some industry jobs opened up after World War I for other women scientists, such as chemists~\cite{rossiter1974}, women physicists were almost entirely dependent on employment at women's colleges through the 1920s~\cite{rossiter1982}.  
As an aside, there were many more opportunities for women physicists to be professionally active following World War II~\cite{howes2015}. \\

Returning back from Italy at the start of World War I, Stone took an opportunity to help establish and become the head of the department of physics at Sweet Briar College in Sweet Briar, Virginia~\cite{addams2019}.  Sweet Briar College was a women's college that had opened only a decade prior in 1906 with just 51 students. Stone chaired the physics department 1916--1923~\cite{creese2000}. \\

In 1923, Stone and her sister opened a new Misses Stone's School for Girls in Washington, DC~\cite{creese2000}.  The school would unfortunately again be foiled by contemporary turmoil and this time was forced to close in 1929 due to the Great Depression~\cite{addams2019}.

%%% SECTION %%%
\vspace{5mm}
\section*{The Unknown Legacy}
\vspace{3mm}

Stone was in her late 90s when she died in 1966 in North Miami, Florida~\cite{addams2019}.  For decades before that, she and her sister had made a living as tutors in Ponce, Puerto Rico~\cite{addams2019}, sustaining her apparent adventurous spirit and love of teaching. \\

Any verifiable photographs or further information about Stone's life are difficult to find, and there is plenty of uncertainty in the information that is available. Some sources incorrectly associate Stone with a contemporary scholar that had the same name who left a request to Phi Beta Kappa in 1934 to establish a graduate fellowship for women~\cite{harvey2000}. Interestingly, \emph{this} Isabelle Stone also got a B.A. from Wellesley College and spent time in Italy, but her Ph.D. work was completed at Cornell University in Greek and Latin studies~\cite{leonard1914,stone1908}.  Other sources incorrectly include a year of study at the University of Berlin as part of Stone's training~\cite{rossiter1982,staley2008}, which instead was carried out by her fellow female co-founder of the American Physical Society, Marcia Keith~\cite{creese2000,howes2015}.  \\
 
Her published work on thin films~\cite{stone1898} opened up a larger field of research for her contemporaries \cite{longden1900,hyatt1912}, and it remains an important contribution to those studying the effects of sample size on charge transport in metallic samples~\cite{munoz2017}.  Her lesser-known role in founding the American Physical Society~\cite{lustig2000} was crucial to creating the national, and later international, community that physicists continue to benefit from today.  Stone's most obscured work as a female physicist traveling to Italy and around the U.S. with a lifelong commitment to educate and inspire young women is an unmeasurable but vital legacy.

\vspace{15pt}

\bibliographystyle{unsrt}
\bibliography{FHP}% Produces the bibliography via BibTeX.

\end{document}